Graphical Abstrad

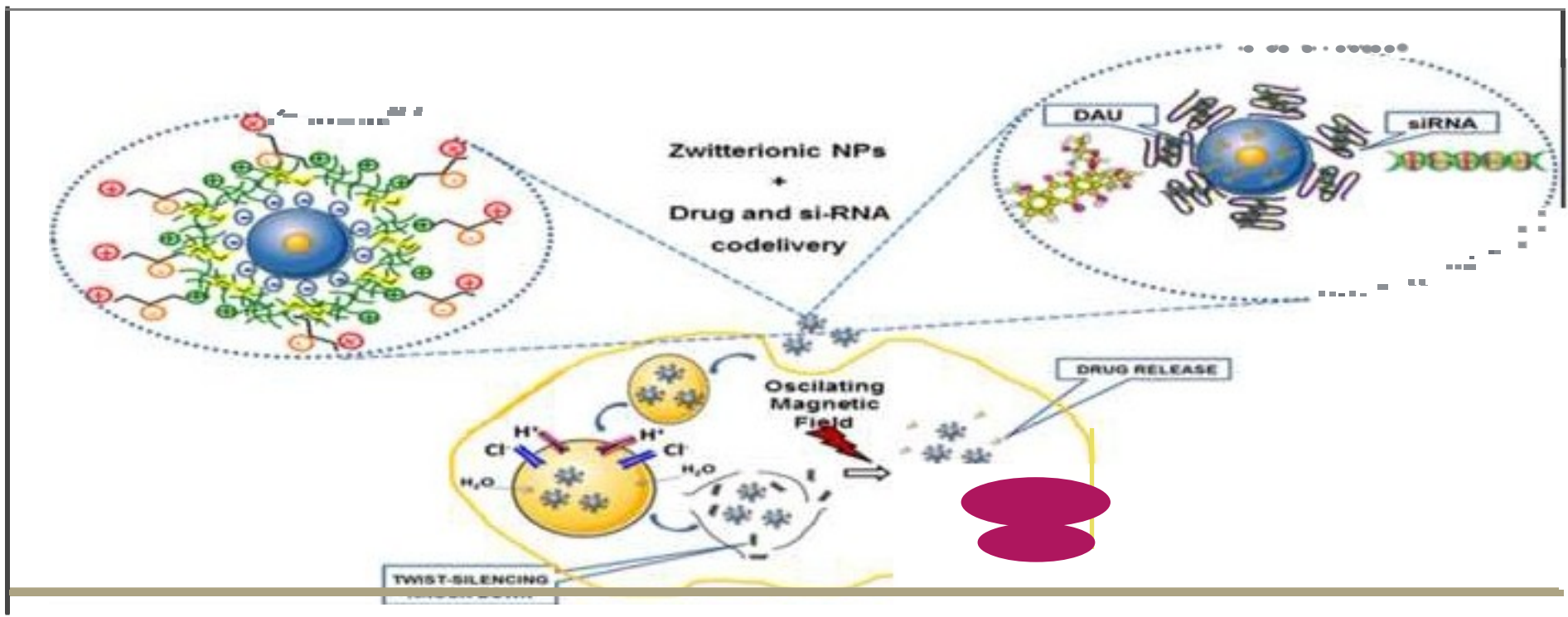

**Highlights**

Multifunctional NPs able to avoid unspecific protein adhesion were prepared.

NPs were crosslinked with GA and decorated with MPC for surface zwitterionization.

Zwitterionic NPs co-delivered siRNA and Dau in response to an OMF has been described.

This approach may provide a therapeutic strategy for the treatment of ovarian cancer.

Manuscript submitted to

# Mesoporous core-shell silica nanoparticles with anti-fouling properties for ovarian cancer therapy

Sandra Sanchez-Salcedo[1,2*], Maria Vallet-Regí[1,2], Sophia Allaf Shanin[3], Carlotta A. Glackin[3], Jeffrey I. Zink[4*]

[1]Department of Inorganic and Bioinorganic Chemistry, Universidad Complutense de Madrid, Hospital 12 de Octubre, Spain.

[2]CIBER-BBN, Spain

[3]Department of Neurosciences, City of Hope, Beckman Research Institute, Duarte, CA, USA

[4]Department of Chemistry and Biochemistry, University of California Los Angeles, Los Angeles, CA, USA.

\* Corresponding authors:

E-mail address: sansanch@ucm.es (S. Sánchez-Salcedo)

E-mail address: zink@chem.ucla.edu  (J. I. Zink)

**Abstract**

Mesoporous silica nanoparticles (MSNPs) have many potential applications in biomedical fields. However, when MSNPs are exposed to plasma, protein adsorption leads to opsonization and decreases blood circulation time. A new multifunctional nanodevice based on polyethylenimine (PEI) coated core-shell $Fe_3O_4@SiO_2$ MSNPs with a zwitterionic 2-methacryloyloxyethyl phosphorylcholine (MPC) surface was designed to minimize unspecific protein adhesion. Particle size measurements demonstrated an excellent non-fouling capacity in solutions containing Bovine Serum Albumin (BSA) and Fetal Bovine Serum (FBS) plasma proteins. The system was used in this study to co-deliver two different cargos: siRNA and daunorubicin. Anti-TWIST siRNA plays critical role in modulating knockdown of TWIST and sensitizing cells to chemotherapeutics such as daunorubicin for ovarian cancer therapy. The drug was released in response to externally controlled oscillating magnetic fields (OMF). siRNA (siGFP) silenced expression of green fluorescence protein (GFP) in Ovcar8 cancer cells, demonstrating the incorporation of core shell MSNPs into cells and siGFP delivery. The synergistic effect of the co-release of anti-TWIST-siRNA loaded in the PEI and daunorubicin loaded in NPs' pores caused increased cytotoxicity in Ovcar8 of up to 50% from both zwitteronic and non-zwitteronic NPs. The system is the first example of silencing by anti-TWITS-siRNA /daunorubicin co-delivered using zwitterionic core-shell nanoparticles with low-fouling adsorption. This engineered multifunctional approach may provide therapeutic potential for the treatment of currently incurable ovarian cancer.

# 1. Introduction

Silica based mesoporous materials and in particular mesoporosus silica nanoparticles (MSNPs) have attracted extensive attention due to their wide spread applications in biomedical fields including biosensing, drug delivery and diagnostic imaging. They are easily chemically modified with biomolecules and are used as theranostic agents for biomedical applications [1-5]. The principal requirements for such nanocarriers include safety (non-toxic) and absence of undesirable side-effects. When nanoparticles (NPs) enter a biological fluid (blood, plasma or interstitial fluid) they are coated with proteins, the "protein corona", that may lead to the exposure of new epitopes, altered function and/or avidity effects [6]. The protein corona spontaneously forms upon exposure to proteins and may consist of multiple layers that have different affinities to the NPs surface, e.g. the soft corona consisting of the external layer of proteins weakly interacting with the NPs, and the hard corona that strongly adheres to the NPs surface [7]. Because the protein corona affects the properties of the NPs surface, it has significant impact on the interaction between the NPs and cell walls [8-10]. Many studies are described in the literature where NP cellular uptake involves the presence of a protein corona. A key factor affecting their final uptake rates is their adhesion to the cell membrane [11-14]. In order to be an effective delivery vehicle, the NPs must have prolonged blood circulation times and must escape the uptake by mononuclear phagocytes, macrophages and the reticuloendothelial systems [12]. The interaction between the positively charged polyethyleneimine (PEI) polymer and the negatively charged proteins might cause a protein corona effect in this type of nanocarriers. Hence, many PEIs have been chemically modified to decrease the side effect and improve their transfection efficiency [15,16]. An interesting approach to avoid this effect is the zwitterionization of MSNPs surface for imparting antifouling properties. Zwitterionic materials are a new class of antifouling surfaces that have demonstrated to be efficient due to the formation of a strong hydration layer through electrostatic interaction. Their weak interactions with serum proteins and their non-toxic nature make them ideal candidates for gene and drug delivery vectors [17-20].

Ovarian carcinoma is a deadly disease because many patients are diagnosed in the metastatic stage of the disease and drug resistance is a major problem [21]. Chemotherapy and surgery are the initial treatment options, but the development of drug resistance is almost universal. For this reason, much effort has been spent on developing alternative therapeutic strategies based on targeted gene therapy in the attempt to exert a cytotoxic action both on the primary tumor and on the peripheral metastases. In recent years, co-delivery of chemotherapy drugs and siRNA knockdown target genes to promote anti-tumor therapy has provided a new approach for cancer treatment and has generated widespread attention [22]. Huang *et al.* reported that administration of polymeric nanoparticles able to deliver diphtheria toxin suicide protein encoding DNA, combined with transcriptional regulation to target gene expression suppresses ovarian tumor growth and reduces tumor burden [20]. Bai *et al*. designed cationic heparin-polyethyleneimine nanogels able to inhibit cell viability by apoptosis induction achieving a tumor weight reduction of~58.55%. [23]. More recently, Robert*s et al.* [24] used the protein TWIST as a therapeutic target. TWIST is a developmental transcription factor reactivated in cancers and linked to angiogenesis, metastasis, cancer stem cell phenotype, and drug resistance. They successfully delivered an anti-TWIST siRNA in combination with cisplatin from PAMAM dendrimers and PEI coated silica NPs. Their results revealed a significant impediment of metastatic growth and significant reduction of the tumor size in animal models of a metastatic and chemoresistant phenotype.

Another key element in addition to delivering siRNA is the ability to release anticancer drugs on demand in the desired location after reduction of drug resistance [25,26]. External control of the release can be achieved by using iron oxide superparamagnetic nanoparticles that generate heat upon exposure to an oscillating magnetic field (OMF) application and uncap pores in MSNs to release the anticancer agent at a specific rate and site, overcoming the problems of conventional techniques for diagnosis and therapy [27]. The core-shell NPs respond to oscillating magnetic fields by generating local nanoparticle heating but the bulk heating effect is negligible, allowing controlled drug delivery strategies [28]. In this work, we present a proof of concept of a new smart multifunctional nanocarrier consisting of core shell silica mesoporous nanoparticles able to avoid unspecific protein adhesion and release two different cargos, anti-TWIST siRNA and daunorubicin, in response to an external oscillating magnetic field.

**2. Experimental Section**

**2.1 Synthesis of Core-shell $Fe_3O_4$@$SiO_2$ mesoporous nanoparticles and PEI coating**

*2.1.1 Synthesis of Core- and functionalization of the surface with DPTES.*

Core-shell $Fe_3O_4$@$SiO_2$ mesoporous nanoparticles were obtained through a [29]. Briefly, 1.25 mg/L of a chloroform dispersion of superparamagnetic iron oxide nanoparticles (SPION, Ocean Nanotech, San Diego, CA) was added to a 22 mL water solution containing 80 mg of CTAB. The mixture was then sonicated for 10 min to allow an homogeneous dispersion of the organic solvent in the water phase after which the resulting solution was stirred at 85 °C to allow the chloroform to evaporate during 10 min. Once the solution became clear the flask was sonicated for 2 min to ensure a good dispersion of the SPION After another 10 minutes at 85 °C, 25 mg of arginine was added and finally 200 μL of TEOS were added drop by drop. In order to obtain functionalised core-shell nanoparticles, 20 μL (2-diethylphosphatoethyl)triethoxysilane (DPTES) (95wt%, Gelest Inc.) was added to the mixture after 3 h leading to nanoparticles with phosphonate group on the surface. The surfactant was removed by ionic exchange soaking nanoparticles in 200 mL of a $NH_4NO_3$ solution (6 g/L) in ethanol at 80 ºC two times during 2 hours under magnetic stirring. MSN were collected by centrifugation, washed three times with ethanol.

*2.1.2 Zwitterionization of the NPs-PEI surface with 2-Methacryloyloxyethyl phosphorylcholine.*

The addition of the cationic PEI coating (1.8 kDa branched polymer, Gelest) to the MSNPs has previously been described [30]. In short, homogeneously suspension of MSNPs was dispersed in PEI ethanolic solution (2.5 mg/mL) which was then sonicated and stirred 30 min (repeated twice). The NPs were then washed in ethanol to remove unbound PEI leading to NPs-PEI. 1 mg/mL of NPs-PEI was mixed in water with 1.2 μL de GA/mg NP glutaraldehyde (GA, 50 wt.%, Sigma-Aldrich, USA). After 30 min. a solution of MPC were added and mixed during 24h at different MPC/DPTES ratios (x:1), where x=0.25, 0.50, 0.75 and 1, giving rise to 25-MPC, 50-MPC, 75-MPC and 100-MPC zwitterionic NPs, respectively. Finally, the samples were washed three times with water giving rise to MPC-25, MPC-50, MPC-75- and MPC-100. NPs were characterized by transmission electron microscopy (TEM), dynamic light scattering (DLS), ζ-potential, nitrogen absorption, termogravimetric analysis (TG), Fourier transformed infrared (FTIR).

*2.1.3 Characterization of NPs-PEI with and without 2-Methacryloyloxyethyl phosphorylcholine.*

The morphological study of MSN was conducted by transmission electron microscopic (TEM) experiments using a FEI Tecnai 12 microscope operating at 120 kV. The $N_2$-adsorption-desorption isotherms were measured with a Micromeritics ASAP 2020 instrument (Micromeritics Co, Norcross, USA).The surface area was calculated by the Brunauer–Emmett–Teller (BET). The pore size distribution was calculated by Barrett–Joyner–Halenda (BJH) method according to the adsorption branch of the isotherm. ζ-potential and dynamic light scattering (DLS) measurements were carried out using a Malvern Zetasizer NanoZS. Samples were prepared at a concentration of 0.1 mg/mL by dispersing the NPs in 10 mM PBS with pH 7.4 at 25ºC. Standard deviations were calculated with 8 runs. FTIR spectra were recorded on a Thermo Nicolet 6700 FTIR system. The sample was prepared as pellets using spectroscopic grade KBr. 1H NMR spectra of the samples were recorded on a Bruker Avance 300 MHz NMR spectrometer using $D_2O$ as the solvent at 25ºC. TG measurements were carried out under a dynamic air atmosphere between 25 and 950ºC with a heating rate of 5ºC/min using a Perkin-Elmer Diamond analyzer (Perkin-Elmer, USA).The magnetic properties of zwitterionic and non-zwitterionic NPs-PEI in a SQUID (MPMS XL-7, Quantum Design) device were carried out at 300 K (Figure S2).

*2.1.4 Antifouling ability evaluation*

The anti-fouling ability of the zwitterionic and non zwitterionic NPs was evaluated in BSA and FBS solutions using dynamic light scattering (DLS) method. Samples were prepared with ultrapure water at a concentration of 0.1 mg/mL by dispersing NPs in a 1 mg/mL BSA and 10% FBS solution using ultrasound at pH 7.4. The distribution of diameters was measured on a Zetasizer (Malvern, UK) before and after incubated in the BSA and FBS solution for 2, 24 and 72 h for NPs-PEI, 25-MPC, 50-MPC, 75-MPC and 100-MPC samples. The mean value was used as hydrodynamic diameter (Dh).

*2.1.5 Loading of core-shell mesoporous nanoparticles with daunorubicin and dye*

Cargo loading was accomplished by soaking the nanoparticles in a saturated solution of Hoechst 33342 or daunorubicin to fill the mesoporous structure by diffusion.

Containment of cargo in the pores was achieved by adding PEI coating which electrostatically binds the molecular thread on the phoshponate silica nanoparticle surface. Then, the nanoparticles were washed thoroughly with water to remove excess dye or drug adsorbed on the silica surface. The addition of the cationic PEI coating to the MSNPs has previously been described [31]. In short, PEI ethanolic solution (2.5 mg/mL) was added to the NPs, then sonicated and stirred for 30 min (repeated twice). The NPs were then washed in ethanol to remove unbound PEI leading to NPs-PEI.

Prior to the release of the drug in the presence of cells, release measurement test was performed on the NPs applying OMF. 1mg/mL of the NPs-PEI, 25-MPC, 50-MPC, 75-MPC, 100-MPC samples are dispersed in PBS, previously loaded with 1 mM Hoechst 33342. The release of the dye was performed at room temperature (RT) and at 50ºC in a water bath between 15 and 360 min. Measurements were performed on a spectrometer (Spectra Pro 2300i, Acton Research Corporation).

The release of the dye with and without applying the external OMF stimulus in frequency range of 500 kHz with amplitude of up to 37.4 $kAm^{-1}$. Before that, the samples were stabilized for 2 h with no stimuli. Samples are placed inside a water-cooled copper coil which produced the OMF. The release profiles of Hoechst 33342 from NPs were monitored at three cycles of 10 min (30 min) and 6 cycles of 10 min (60 min).

**2.2 Cell culture test**

**2.2.1 Cell culture**

Ovcar8 were obtained from ATCC and were grown in RPMI 1640 (Genesee Scientific, San Diego, CA) in a tissue culture incubator at 37 °C, 5% $CO_2$, and 90% humidity. Growth medium was supplemented with 10% fetal bovine serum and 1% penicillin/streptomycin. Cells were passaged every 2-4 days using 0.25% trypsin (Genesee Scientific). Cells were passaged every 2-4 days using 0.25% trypsin (Genesee Scientific). To allow for imaging of cells, we created a line of Ovcar8 that stably expressed GFP and firefly luciferase (ffluc). These cells were created with the aid of a CMV lentiviral construct that encodes a fusion protein of GFP and ffluc separated by a three glycine linker. This stable cell line was used for all experiments.

### 2.2.2. Fluorescence microscopy

We determined the time it takes NPs are internalized in the cells they load the NPs with a siRNA marked with Alexa647. For this purpose $2.5 \cdot 10^4$ cells / ml were seeded in a 6-well plate the day prior to incorporation of the NPs. At the same time, 70 ul of the nanoparticles were loaded into 10μL of siRNA-Alexa647 (Qiagen, Valencia, CA) and 20μL of free RNAases water (Fisher Scientific, USA) were washed tree times with PBS at 12 hours times and fixed by 70% ethanol for 15 min. The cells were observed with a fluorescence microscope (Nikon, Nikon Instruments Inc.). Cells in green (450/510 excitation/emission) and red NPs loaded with siRNA-Alexa647 (650/665 of excitation/emission).

### 2.2.3 eGFP gene silencing and transfection

Ovcar8 cells were seeded in 6-well plates at a seeding density of $3.5 \cdot 10^3$ cells/mL 24 h prior to the experiment. In parallel 70 μL of nanoparticles were loaded with 10μL siRNA or (Dharmacon™ Accell™ eGFP Control siRNA, USA) and 20μL RNAase free water (Fisher Scientific, USA) and mixed for 15 hours. Then, 100 μL of NPs-siGFP (siRNA against eGFP) were added to cells and incubated at 37 °C and 5% $CO_2$ during 1, 2, 3 and 4d. The medium was changed 24 h of the addition of the NPs and every two days of the trial. The cells with siGFP were observed in a fluorescence microscope (Nikon, Nikon Instruments Inc.) with a wavelength of 410 nm excitation and emission of 510 nm.

**2.2.4 Cargo and release of daunorubicin: Cell viability test.**

Ovcar8 cells were seeded in 8-well plates at a seeding density of $3.5 \cdot 10^3$ cells/mL 24 h prior to the experiment. In parallel, 17.5 μL of concentration C1 (17.5ug/mL NPs) and 35 μL of concentration C2 (35ug/mL NPs) were prepared. 10 mg NPs were previously loaded with 2 mL of daunorubicin in ethanol (1 mM) in dark overnight, washing three times followed of PEI and/or MPC functionalization as previously described. The amount of loaded fluorescein was 8±2%, determined by TG. The culture medium was changed after 24h of incubation and three cycles of 10 min (30 min) and 6 cycles of 10 min (60 min) applied to the cells of an oscillating magnetic field (375 kHz, 20 kA m$^{-1}$) at RT.

**2.2.5 Zwitterionic and non zwitterionic NPs with anti-TWIST-siRNA (si419): Cell viability test.**

Previously published siRNA sequences against TWIST (si419-passenger, 5'-GACAAGCUGAGCAAGAUU-3'; si419-guide, 5'-AAUCUUGCUCAGCUUGUCCUU-3'; si494-passenger, 5'-GCGACGAGCUGGACUCCAA-3'; si494-guide, 5'-UUGGAGUCCAGCUCGUCGCUU-3') were used [30]. Two chemical modifications (addition of 2′-O-methyl and inverted abasic ribose) were made to the si419 passenger/sense strand for all experiments. No chemical modifications were made to the si494 sequences. siRNA duplexes were formed by mixing equal molar volumes of each strand together and heating them in a hot block (100 °C) for 10 min, then removing the block from the heat source to cool to RT. Negative control siRNA (siQ, labeled with AlexaFluor®647) was All Stars Negative Control siRNA from Qiagen (Valencia, CA). There were assay five different conditions with NPs-PEI, 50-MPC and 75-MPC: control NPs (NPs), siRNA loaded inNPs as negative control (siQ), NPs loaded with anti-TWIST-siRNA (Si4199 and daunorubicin without OMF (Si419-Dau), NPs loaded with Si419 and daunorubicin with OMF (Si419-Dau*) and two dosis of NPs loaded with Si419 the first one and NPs loaded with daunorubicin the second one with OMF with the same final NPs amount (Si419-Dau*-TS). Transfection of siRNA with zwitteronic and non zwitterionic NPs was carried out by incubating the NPs with the siRNA overnight at 4°C while rotating the tube constantly. The mixture consisted of a similar protocol of eGFP gene silencing with 35 μL diluted from 0.5 mg/mL and using Si419 diluted in free RNAase water mixed during 15h. The final concentrations of NPs and siRNA applied to cells were 35 ng/μl and 100 nM, respectively. Following Roberts CM et al. protocol [24] NPs were added to cells and were incubated at 37ºC and 5% $CO_2$ during 7d. The media were changed after 24 h of incoporating zwitterionic and non zwitterionic NPs and each three days of assay. A second dosis of Dau charged NPs were added after 5d of assay in Si-TWIST-Dau* plate. OMF were applied in Si-TWIST-Dau* and Si-TWIST-Dau* (two steps) samples. The plate was placed inside a water-cooled copper coil which produced an OMF with the same above conditions. The culture medium was changed after 24 of incubation and three cycles of 10 min (30 min) and 6 cycles of 10 min (60 min) applied to the cells of an OMF (375 kHz, 20 kA m−1) at RT. Mitochondrial activity of cells in contact with NPs was measured using the 3-[4.5-dimethylthiazol-2-yl]-2,5-diphenyltetrazolium bromide (MTT) test. The MTT assay is based on the activity of mitochondrial dehydrogenases of living cells that reduces yellow MTT to blue formazan, and the colored solution is measured spectrophotometrically at 570 nm. Following this incubation period media was removed from each well and 110 μl MTT diluted in complete media (0.45 mg/mL) was added. Incubation lasted for 3 hours at 37°C, 5% $CO_2$, with 90% humidity. Following the incubation period the MTT media was removed and 110 μl DMSO was added to each well and the plate was gentle shaken for 15 min. Statistical comparisons were made by analysis of variance (ANOVA). The Scheffé test was used for post hoc evaluation of differences between groups. In all statistical evaluations, p < 0.05 was considered as statistically significant.

## 3. Results and discussion

### 3.1 Characterization of core shell $Fe_3O_4@SiO_2$ MSNPs

The synthetic protocol is shown in Figure 1. Core-shell $Fe_3O_4@SiO_2$ mesoporous nanoparticles functionalized with DPTES (negatively charged), were grafted with the cationic polymer PEI (1.8 kDa branched polymer), to ensure siRNA in MSNPs [32]. Subsequently, the PEI was crosslinked with GA give rise to an imine Schiff base, avoiding degradation of $Fe_3O_4@SiO_2$ NPs-PEI in the next step in water media (Figure S1). Finally, functionalization of amine from PEI by Michael addition reaction with biomimetic MPC monomer was performed [33]. The partial functionalization of PEI give rise to NPs with *zwitterionic fragments* named as 25-MPC, 50-MPC, 75-MPC and 100-MPC *zwitterionic NPs*. TEM images show monodispersed NPs before and after covering with PEI and MPC functionalization (Figure 2A). The zwitterionic and non zwitterionic NPs exhibit a ~ 70 nm diameter with a $Fe_3O_4$ core of 20 nm. DLS analysis showed an average hydrodynamic size of a ~74 nm and ~125 nm without and with PEI in PBS, respectively. Bare NPs exhibits a negative ζ-potential (-31.2 mV) and after PEI coating resulted in particles with a positive ζ-potential of +24.5 mV in PBS.

Textural properties of the MSNPs have been performed by $N_2$ adsorption/desorption analyses (Figure 2B). Similar behavior was obtained in all NPs (Table 1). The isotherms of NPs-PEI, MPC-100 and calcined MPC-100 samples were plotted as a representative example. They can be identified as type IV according to the IUPAC classification, which are typical of mesoporous solids [34]. The presence of H1 type hysteresis loops in the mesopore range indicates the existence of open ended cylindrical mesopores with narrow pore size distributions [35, 36]. The surface area ($S_{BET}$) experienced an increase when the samples are calcined at 600ºC indicating the previous coverage and after calcination, the maintenance of mesoporosity after the PEI and MPC functionalization for MPC-100. A spherical silica structure with a radial orientation and open pores about ~1-2 nm in diameter were obtained. Pore volume, pore diameter and surface area displayed a progressive decrease from NPs functionalized with PEI and for all zwitterionic NPs which is in agreement with the appropriate successively functionalization of NPs and the increase of $S_{BET}$ when the samples are calcined indicating the maintenance of mesoporosity, as was expected (Table 1).

The chemical structures of NPs-PEI with and without MPC were identified with Fourier transformation infrared spectroscopy (FTIR) and the results are shown in Figure 2C. FTIR spectra for the bare NPs exhibiting bands of Si–O at 1050, 802 and 470 cm$^{-1}$. Characteristic peaks of PEI at 3272 cm$^{-1}$ (–N–H stretching), 2940–2830 cm$^{-1}$ (–C–H stretching), 1576 cm$^{-1}$ (–N–H bending), 1465 cm$^{-1}$ (–C–H bending) and 1350–1000 cm$^{-1}$ (–C–N stretching). Unlike PEI, the FITR spectrum of polyamine microspheres displays a distinct peak at 1650 cm$^{-1}$, which is the stretching band of –C=N, indicating the Schiff reaction between the amine groups of PEI and the aldehyde groups of glutaraldehyde. It should be noted that the presence of water molecules also result in the appearance of an absorption peak at 3424 cm$^{-1}$ [37]. Zwitterionic NPs exhibiting bands at 1718, 1242, 1078 and 970 cm$^{-1}$ corresponding to (C=O), (O=P–O-), (P–O–C) and (–N+-(CH$_3$)$_3$–) respectively, indicating the functionalization of NPs with MPC. Moreover, no signals at 1321 cm$^{-1}$ (C=C), 1298 cm$^{-1}$ (C=C) indicate the absence of methacrylic C=C double bonds so the anchoring (not adsorption) of MPC can be confirmed. [38, 39]. The 1H NMR spectra for MPC product, zwitterionic and non zwitterionic samples are shown in Fig. 2D. Signals between 2.5 and 3.2 ppm correspond to PEI polymer. Signals at 1.89, 5.77 and 6.15 relative to C=C bond of MPC product indicating an anchored to PEI-NPs. MPC 1H-NMR (300 MHz in D$_2$O): $d$ (ppm) =1.89 (s; CH$_2$=CHCH$_3$ 6H), 3.1 (s; –N$^+$(CH$_3$)$_2$– 6H), 3.74 (m; –CH$_2$– 2H), 3.86 (m; –CH$_2$– 2H), 4.15 (m; –CH$_2$– 2H), 4.32 (m; –CH$_2$– 2H), 4.38 (m; –CH$_2$– 2H), 4.65 (m; –CH$_2$– 2H), 5.74 (m; –CH=CHCH$_3$ 1H), 5.77 (m; –CH=CHCH$_3$ 1H), 6.16 (m; –CH=CHCH$_3$ 1H). The amount of PEI and MPC anchored in the surface of NPs were determined by TG analyses. The results reveal a 23.8 % of PEI in NPs-PEI, 0.5% of GA and a progressive increase of MPC for 25-MPC, 50-MPC, 75-MPC and MPC-100 samples (Table 1). In addition, the results obtained by TG measurements revealed that there were different weight losses between materials before and after MPC functionalization. Thus, the successful immobilization of MPC on the nanoparticles has been confirmed.

### 3.2. Antifouling ability evaluation

It is known that proteins adsorbed onto the surface of biomaterials can trigger thrombosis and blood coagulation, leading to life-threatening situations or causing functional failure when biomaterials are used in artificial organs, blood vessels, and other medical devices in contact with blood. This undesirable interaction with non-specific protein adsorption is among the first events that occur at the interface between carrier and human blood. [40] To have a better understanding on how the amino and phsosphate groups from MPC will affect anti-fouling properties of NPs-PEI, their anti-fouling property was evaluated in protein solutions using DLS measurements.

First of all, the stability of the samples in PBS was tested. All the NPs are stable up to 3 days, more than enough time to be internalized by the cancer cells after being administered (Figure 3). After protein non-specific adsorption occurs, the hydrodynamic diameter of NPs will increase. A single component Bovine Serum Albumin (BSA) solution was first used because albumin is the most abundant protein in serum. In addition, to mimic the complex biological system, NPs were incubated in the fetal bovine serum (FBS) solution and the DLS was measured. Before carrying out the BSA study, the ζ-potential of samples were measured in water to estimate the total relative charge of zwitterionic and non zwitterionic NPs-PEI in PBS. The ζ-potentials were positive for all of the NPs, as expected because of the presence of secondary and primary amino groups from the PEI polymer. Moreover, all MPC functionalized NPs presented a positive charge due to their zwitterionic nature and the PEI amino groups. 75-MPC and 100-MCP samples exhibited a slight ζ-potential decrease due to a higher zwitterionization (Figure 3 and Table1). The ζ-potentials of BSA and FBS were also measured and had a negative charge, indicating that an electrostatic attraction to PEI-derivatized NPs should be favourable. In DLS particle sizes remained stable over time in the 50-MPC, 75-MPC and 100-MCP NPs compared to NPs-PEI and 25-MPC samples over time. In NPs-PEI and 25-MPC samples the size increased over time to 340 and 220 nm in FBS solution and up to 225 and 200 nm respectively in BSA solution. In BSA their size is less than 130 nm after 72h. 50-MPC, 75-MPC zwitterionic NPs sizes were maintained at 170 nm in FBS solution (Figure 3). Thus the most suitable samples to avoid nonspecific protein adhesion and thus would avoid opsonization of NPs in the bloodstream were 50-MPC and 75-MPC. This phenomenon can probably be attributed to the strong surface hydration layer formed by the neutral zwitterionic structure. These results indicate protein corona formation around 50-MPC, 75-MPC is inhibited in BSA and FBS media.

### 3.3 In vitro cargo release experiments

The application of an oscillating magnetic field (OMF) causes heating of the iron oxide nanocrystal's core. Thus, prior to studying the release of the drug in the presence of cells, we performed a series of release measurements on NPs as a function of temperature. For that study, 1 mg/mL of the NPs-PEI, 25-MPC, 50-MPC, 75-MPC, 100-MPC in PBS samples are dispersed. The samples were washed multiple times to remove dye from the surfaces until the washings were free of dye. The samples were washed multiple times to remove dye from the surfaces until the washings were free of dye. The measurements of the Hoechst dye release were made between 15 and 360 min at RT and with external stimulus. It is observed that in samples with GA crosslinked PEI and zwitterionization with MPC the release of Hoechst to the medium is faster from zwitterionic NPs when the temperature is applied (Figure 4) as it will be explained later. Dye release data showed in Figure 4A can be fitted to a first-order kinetic model with an empirical nonideality factor ($\delta$) (eq 1) [41]:

$$Y = A(1-e^{kt})^{\delta}$$

where Y is the percentage of dye released at time t, A the maximum amount of dye released (in percentage), and k, the release rate constant. The values for $\delta$ are between 1 for materials that follow first-order kinetics, and 0, for materials that release the loaded drug in the very initial time of analysis. The kinetic fitting indicates that, while the maximum amount of Hoechst released is practically the totality of the loaded dye after 360 min of temperature application, most of the cargo is retained without the external stimulus (>62%) for 50-MPC and 75-MPC zwitterionic NPs. Since $\delta$ gives an approximation of the degree of fidelity of the proposed model to the theoretical first-order kinetics, when $\delta$ is smaller the first-order is less accurate and the delivery of the drugs from mesopores shows a larger burst effect. This latter effect is usually attributed to the adsorption of drug molecules on the outermost surface of the mesopores. As it can be seen in Figure 4B, the values of $\delta$ are larger for Hoechst release at RT than at 50ºC. In addition, the maximum percentage of delivered from NPs (Y), goes up to 83-89%, reached in the first 6 hours with the external stimulus from MPC-50 and MPC-75, respectively. The reduction of the drug release amount in NPs-PEI is attributed to the partial retention of drug molecules onto the mesopores due to higher electrostatic repulsions between cargo and positively charged PEI as we explain immediately above.

**3.4 In vitro cargo release assays by OMF**

The release profiles caused by magnetic heating were studied by monitoring the fluorescence of the released dye from aliquots of the solution taken at different time points (Figure 5). The cuvette containing the particles was placed inside a 5-turn coil and an oscillating magnetic field (375 kHz, 20 kA m$^{-1}$) was applied. The first three points in the graph correspond to equilibration before magnetic activation, showing minimal dye leakage. After 2 h, three cycles of 10 min (30 min) and 6 cycles of 10 min (60 min) and 1 min/cycle heating off were applied; the cooling cycle was used to minimize overheating the bulk solution, monitoring the temperature each 10 min (not higher than 37ºC) (Figure 5A). For therapeutic applications, it is important to take into account that, althougt the system have no specific functionalization for OMF stimuli, the GA crosslinking and MPC anchored to PEI reveal a different behaviour between NPs-PEI and zwitterionic MPC-50 and MPC-75 NPs. Albeit, when OMF stimulus is applied at 30 and 60 min. significatively increase of cargo is released from MPC-50 and MPC-75 NPs. The release after 60 min of OMF cycle was 14% and 23% higher in MPC-50 and MPC-75 NPs compared to NPs-PEI (Figure 5B). Thus, during the OMF stimulus cycles a higher amount of dye release from zwitterionic NPs. Similar behaviour was observed when temperature stimulus is applied to NPs in Figure 4. The daunorubicin is an anthracyclines as doxorubicin. They have the same amino sugar moiety daunosamine, with pK of 8.4 so it is positively charged at physiological pH as well as the cationic dye, Hoechst 33342 (pKa = 11.9). Therefore, these cargos are attracted to negatively charged surfaces where both agents could be released by dropping the environmental pH. Thus, the load of cargo in negatively charged phsophonate NPs (pKa = 2) was enhanced compared to silica NPs (pKa = 7), which is near to its isoelectric point and therefore not quite as effective for electrostatic binding [42]. The electrostatic repulsion with cationic PEI provoke a slower release from NPs-PEI compared to zwitterionic NPs. Since hydrophobic interactions increase and positively repulsions decresase when PEI was cross-linked with GA and zwitterionizated in case of MPC-50 and MPC-75 NPs, so the release of cargo is significantly higher after OMF stimulus was applied. *Gawde et al.* observed a similar behaviour when a higher drug release rate in glutaraldehyde-crosslinked albumin bio-conjugate nanoparticles system. [43]. *Dmour et al.* [44] investigated the potential to stabilize chitosan based NPs prepared by ionotropic gelation method by covalent crosslinking using a dimethylaminopropyl-carbodiimide as a coupling agent and observed that the release of doxorubicin was enhanced upon loading in covalent crosslinked NPs compared to free doxorubicin.

**3.5 Internalization assay**

In vitro cellular uptake experiments were performed using Ovcar8 cells and zwitterionic and non zwitterionic NPs (Figure 6A). To identify the final fate of the internalized nanoparticles, a fluorescence imaging study was carried out by fluorescence microscopy. The uptake of the NPs was evaluated after 12 h cell culture, and is shown in Figure 6A. As has been demonstrated, the cationic polymer PEI could facilitate nanoparticle uptake in some extent, so NPs-PEI is a control in the test [45]. Red Alexa647 labelled siRNA loaded in the NPs and the eGFP transgene green fluorescence properties in Ovcar8 cells was chosen for our internalization in vitro experiments. The orange spots derived from localized nanoparticles (red) in endolysosomes (green) were noticeable for zwitterionic and non zwitterionic NPs. Thus, labelled siRNA loaded in NPs remain on the NPs and the NPs were internalized after 12 h of exposure.

### 3.6 eGFP gene silencing and transfection

It was of great importance to provide evidence of activity of the siRNA inside cells and determine the efficacy of zwitterionic NPs as siRNA carriers. The eGFP transgene was chosen for our initial in vitro experiments because eGFP's fluorescence properties facilitated analysis of silencing efficiency [31].

To visualize the transfection effect, the eGFP expression in Ovcar8 cells was observed using fluorescence microscopy (Nikon). Figure 6B shows some typical fluorescence images of ovarian cancer cells transfected by NPs-PEI, MPC-50 and MPC-75. As can be seen all NPs loaded with siRNA against eGFP gene caused a dramatic decrease in the number of fluorescent Ovcar8 cells after 4 days compared to 1 day. These results demonstrate that both zwitterionic and non zwitterionic nanocarriers are able of sequestering and protecting siRNA, carrying it into cells and silencing the target gene expression of the exogenous green fluorescent protein of ovarian cancer cells.

### 3.7 Cell viability evaluation

### 3.7.1 In vitro cytotoxicity assay.

First, Ovcar8 cells were seeded in 8-well Chamber Slide™ system (Nunc® Lab-Tek) at a seeding density of $3.5 \cdot 10^3$ cells/well 24 h prior to the experiment. We assessed the cytotoxicity of NPs-PEI, MPC-50 and MPC-75NPs with and without daunorubicin (Dau) at 17.5 ug/mL (1) and 35 ug/mL (2) of NPs applying 30 and 60 min. of OMF external stimulus (Figure 7A). In presence of the OMF, the local heating caused by the magnetic NPs, facilitated the release of Dau from the silica pores and thus decreases proliferation of ovarian cancer cells. The cytotoxicity was similar for 30 min and 60 min of OMF cycles in NPs-PEI and MPC-50. MPC-75 NPs exhibit significantly higher cytotoxicity behaviour at 35 ug/mL NPs with 60 min of OMF and compared to NPs-PEI and MPC-50, due to the release of cargo in this system was higher after two cycles of OMF (Figure 7A). Thus, the zwitterionization of NPs exhibited at least similar cytotoxicity of non zwitterionic NPs-PEI with *c.a.* 20 and 30 % of cell killing. The effect of the OMF on the cells was examined without drug loading under the same conditions as a control. When these NPs were endocytosed into the cells and exposed to the OMF, no significant cytotoxicity was observed.

### 3.7.2 Co-delivery efficacy of the siRNA and Daunorubicin loaded NPs in vitro

Since we demonstrated a higher or similar killing cancer cells at 60 min of OMF and 35 µg/mL of zwitterionic and non zwitterionic NPs, we tested the toxicity of co-delivery of anti-TWIST siRNA (Si419) and Dau from the NPs in these conditions (Figure 7B). To evaluate the effects of both Si419/Dau co-delivery and an active targeting strategy on the cytotoxicity in Ovcar8 cells, the cytotoxicities of the zwitterionic and non zwitterionic NPs loaded with both Si419/Dau and applying or not OMF was measured by MTT assay after 7d (Figure 7B). NPs-PEI, MPC-50 and MPC-75 in Si419-Dau and Si419- Dau* conditions induced a similar decrease in cell viability in Ovcar8 cells and significantly higher cytotoxicity compared to controls of NPs and siQ conditions. These NPs and siQ conditions should not exercise any cytotoxic effect due to they are negative controls. Moreover, MPC-50 and MPC-75 NPs with Si419-Dau* conditions all markedly decreased up to 50% the cell viability compared to controls, indicating a significant synergetic effectiveness for the combination of siRNA and Dau. Thus, the siRNA/Dau combined treatment delivered with OMF by zwitterionic MPC-50 and MPC-75 NPs produced significantly higher cytotoxicity than the same with no OMF (Si419-Dau) and the system in two steps (Si419-Dau*-TS). This system was not so effective since the two separate doses (17.5µL NPs/dose) with five days of difference did not have the same killing effect than the synergic Si419-Dau* conditions. Furthermore, the cytototoxic effect was up to 20% higher compared to zwitterionic and non zwitterionic NPs loaded only with Dau. The proposed mechanism of the NPs activation could be mediated by the increased cellular accumulation due to the potential of PEI to enter the lysosomal compartment. The 'proton sponge' nature of PEI is thought to lead to buffering inside endosomes. Nanoparticles coated with polycationic polymers can deliver anticancer drugs and/or DNA/siRNA into the cytosol, based on the endosomal escape mechanism to achieve tumor static effects. The additional pumping of protons into the endosome, along with the concurrent influx of chloride ions to maintain charge neutrality, increases ionic strength inside the endosome. This effect then causes osmotic swelling and physical rupture of the endosome, resulting in the escape of the vector from the degradative lysosomal trafficking pathway [6]. It is worthy to note that zwitterionic NPs maintained that properties after MPC functionalization. These results are agree with the findings of several authors that have demonstrated better transfection efficiency modifying low molecular weight PEI with a hydrazone-based crosslinker [46-48].

In summary, these NPs have been designed for a multiple function as co-delivery of chemotherapy drug and metastasis related gene silencing siRNA to enhance cancer therapy (Figure 8). This proof-of-concept of proposed multifunctional nanocarriers designed to kill ovarian cancer cells substantially enhances efficiency of chemotherapy with low-fouling properties.

## 4. CONCLUSION

In this work, a proof of concept of a new smart multifunctional nanoparticles able to decrease unspecific protein adhesion and release two different cargos, siRNA and daunorubicin, in response to external stimuli, such as oscillating magnetic field, has been presented. To our knowledge, this research represents the first example of silencing by siRNA co-delived with Dau using zwitterionic core-shell nanoparticles with resistance to nonspecific protein adsorption. Thus, this combinatorial approach may provide therapeutic potential for the treatment of ovarian cancer to translate these in vitro results to clinical applications.

**Acknowledgements**

This study was supported by research grants from the Ministerio de Economía y Competitividad (CAS15/00153). The work at UCLA was supported by grant NIH R01Ca133677 (F. Tamanoi and J.I.Zink). M. Vallet-Regí acknowledges funding from the European Research Council (Advanced Grant VERDI; ERC-2015-AdG Proposal No. 694160). We thank Professor F. Tamanoi, Department of Microbiology Immunology and Molecular Genetics, UCLA, Los Angeles, CA, USA for laboratory support.

**Figure Captions**

**Figure 1.** Synthesis scheme of zwitterionization of NPs-PEI**.**

**Figure 2. A.** Characterization by Dinamic Light Scattering and ζ-potential of NPs and NPs-PEI. **B**. $N_2$ adsorption isotherms surface area and pore size of NPs and NPS-PEI, 100-MPC-calcined and 100-MPC. **C**. FTIR bands zwitterionic and non-zwitterionic of MSNPs **D**. H NMR spectrum of PEI-MPC with varied graft ratios (in $D_2O$).

**Figure 3.** Evolution of hydrodinamic diameter of 0.1mg/mL of NPs-PEI, 25-MPC,50-MPC,75-MPC and 100-MPC with soaking in 1 mg/mL BSA and 10% FBS solution. Evolution of hydrodinamic diameter of 1mg/mL of all nanoparticles in PBS.

**Figure 4. (A)** Hoechst 33342 release from NPs-PEI, 50-MPC and 75-MPC in PBS solution versus time with external stimuli temperature. **(B)** Kinetic model of cargo release from zwitterinic and non zwitterionic NPs at room temperature and 50ºC.

**Figure 5.** Hoechst 33342 release from NPs-PEI, 50-MPC and 75-MPC in PBS solution versus time with external stimuli oscilating magnetic field (OMF) **(A)**. Samples are placed inside a water-cooled copper coil which produced an oscilating magnetic field (OMF) in frequency range of 500 kHz with an amplitude of up to 37.4 $kAm^{-1}$ after 2 h with no stimuli. **(B)** Release profiles of Hoechst 33342 from NPs at 30 and 60 min of OMF.

**Figure 6. (A)** Fluorescence microscopy images of EGFP expressing Ovcar8 cells incubated with NPs loaded with Si-RNA-labeled with Alexa647. From left to rigth: Green fluorescence of EGFP-expressing Ovcar8 cells; Red fluorescence (Alexa647-siRNA-NPs) of internalized NPs and overlay images of the two fluorescence channels after 12 h. **(B)** Images of NPs-PEI, 50-MPC and 75-MPC mediated eGFP silencing with fluorescence microscopy. Untreated Ovcar-8 cells as a negative control (control) and eGFP-expressing Ovcar8 cells with NPs after 1day and 4 days.

**Figure 7.** Results of Ovcar-8 cells exposure to Si-RNA and the OMF. **Panel (A)** shows fluorescent microscope images of cells after 60 min of OMF without NPs and with 35 µg/mL of NPs-PEI, 50-MPC and 75-MPC. NPs without Si-RNA and daunorrubicin (DAU) no cytotoxicity occurred in cells. After 30 and 60 min of OMF with 35 µg/mL the citotoxicity of NPs were higher compared to 17.5 µg/mL with *c.a.* 20 and 30 % of cell killing. **Panel (B)** Comparison of the cytotoxic effects of different co-delivery modalities of Si-RNA and/or Dau in Ovcar-8 cells by MTT assay. The Ovcar-8 cells were treated by 35 µg/mLNPs, siRNA-NPs, siRNA-DAU-NPs, siRNA-DAU-NPs* with OMF and siRNA-DAU-NPs, siRNA-DAU-NPs* in two steps, first dose of siRNA- NPs and afer 5 days second dose of DAU-NPs for 7 days of assy. Data were expressed as the mean ± SEM of three independent experiments. *$^{\neq}$p > 0.05.

**Figure 8.** Shematic of the proton sponge effect leading to lysosomal damage and induction of citytoxitity with co-delivery siRNA and daunorubicin by multifunctional zwitterionic nanoparticles.

**Table**

| Sample | $S_{BET}$ (m$^2$/g) | $V_T$ (cm$^3$/g) | $D_P$ (nm) | % PEI (TG) | % MPC (TG) |
|---|---|---|---|---|---|
| **NPs-PEI(CAL)** | 964 | 1.12 | 2.5 | -- | -- |
| **NPs-PEI** | 250 | 0.60 | 1.2 | 23.8 | -- |
| **MPC-25** | 205 | 0.58 | 0.9 | 23.8 | 3.5 |
| **MPC-25(CAL)** | 640 | 0.87 | 2.1 | -- | -- |
| **MPC-50** | 204 | 0.55 | 0.9 | 23.8 | 7 |
| **MPC-50(CAL)** | 630 | 0.85 | 1.8 | -- | -- |
| **MPC-75** | 191 | 0.54 | 0.9 | 23.8 | 12 |
| **MPC-75(CAL)** | 652 | 1.1 | 1.8 | -- | -- |
| **MPC-100** | 194 | 0.47 | 1.1 | 23.8 | 16 |
| **MPC-100(CAL)** | 638 | 0.90 | 1.9 | -- | -- |

**Table 1.** Textural properties derived from N$_2$ sorption measurements for bare MSNPs, NPs-PEI and MPC NPs before calcination and calcinated (CAL).

S$_{BET}$ is the total surface area determined by the BET method between the relative pressures (P/P$^0$) 0.05-0.25. V$_T$ is the total pore volume obtained using the t-plot method. D$_P$ is the pore diameter calculated by means of the Barrett-Joyner-Halenda (BJH) method from the adsorption branch of N$_2$ isotherm. The percentage of functionalization with PEI and later MPC incorporation was estimated from thermogravimetric (TG) measurements.

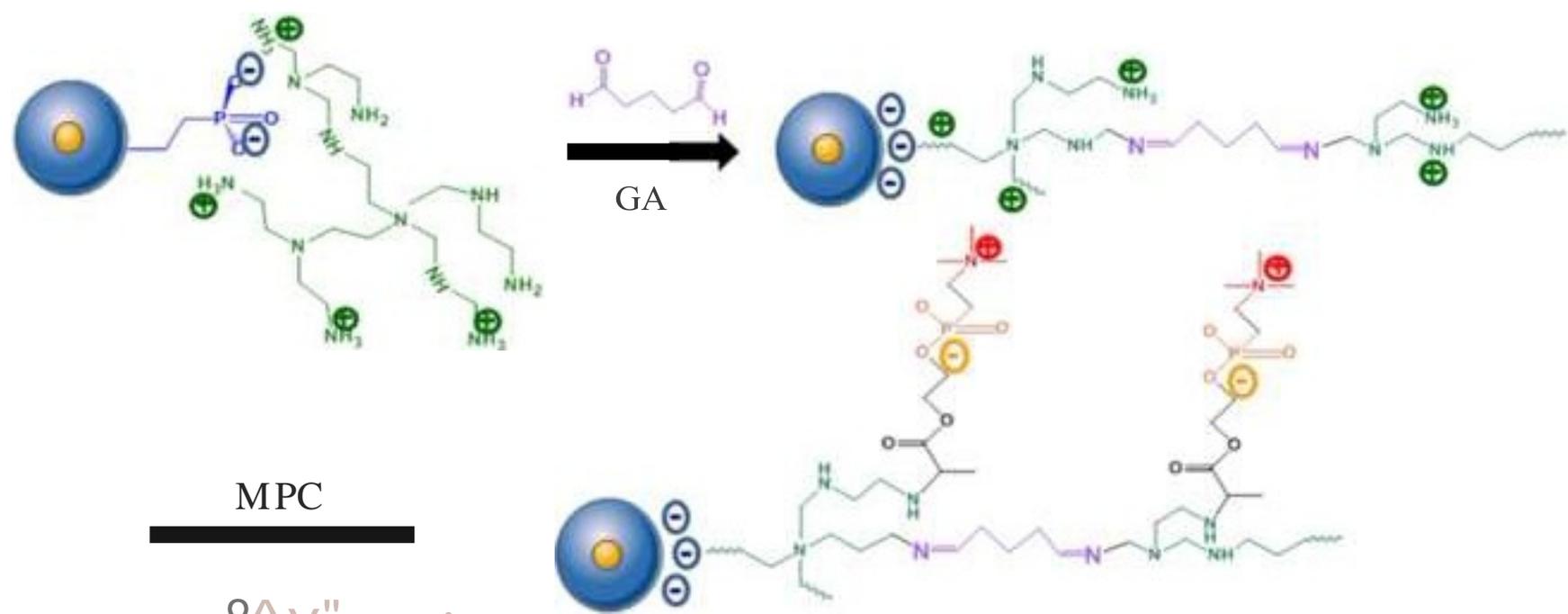
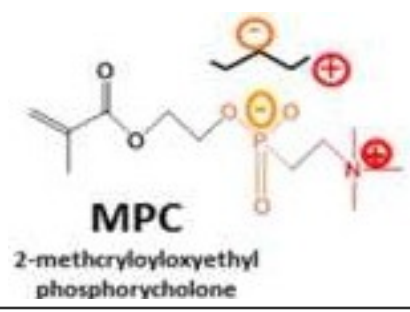
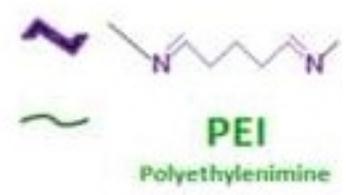
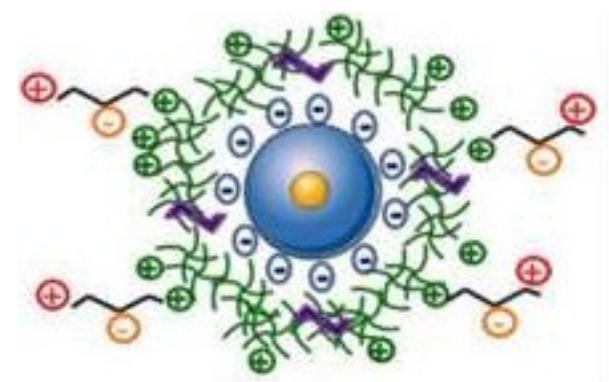

Figure 2

A
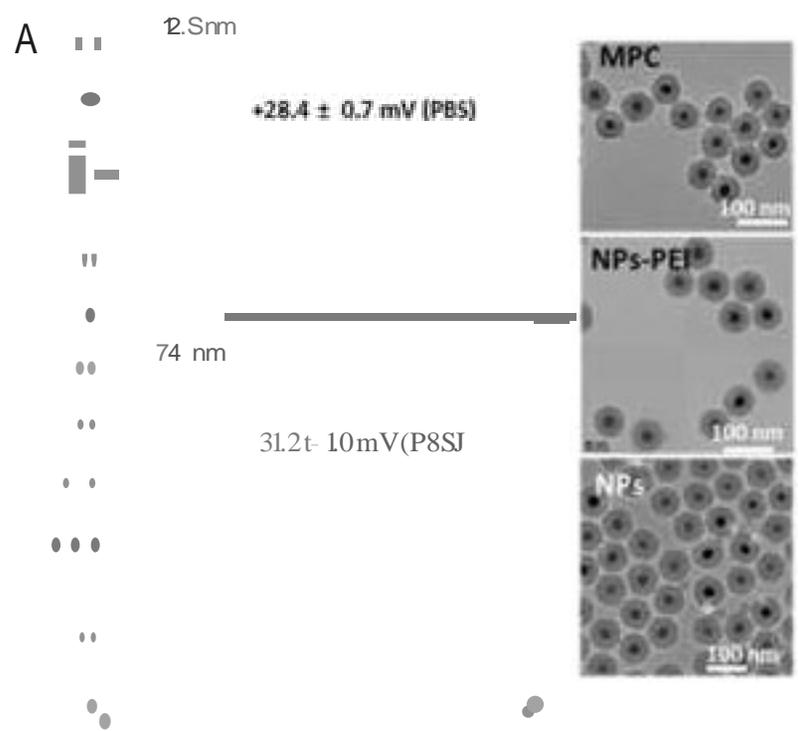

e
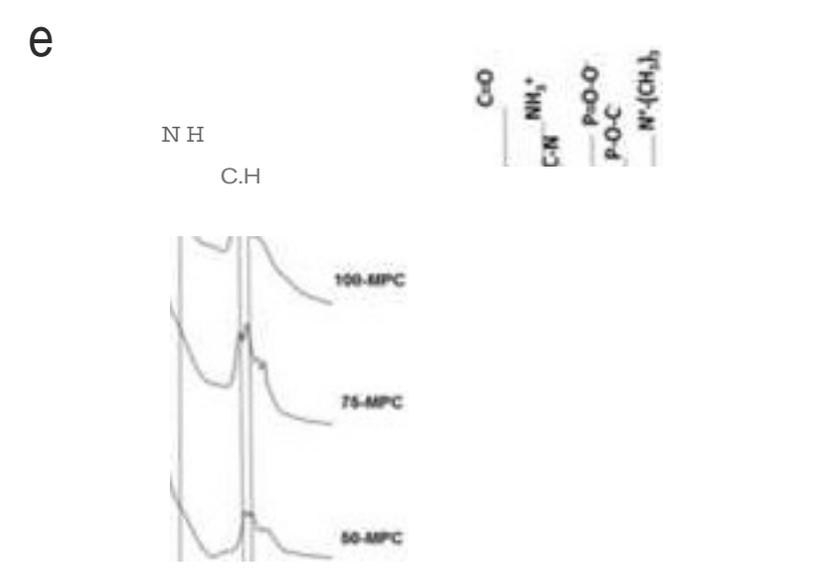

B
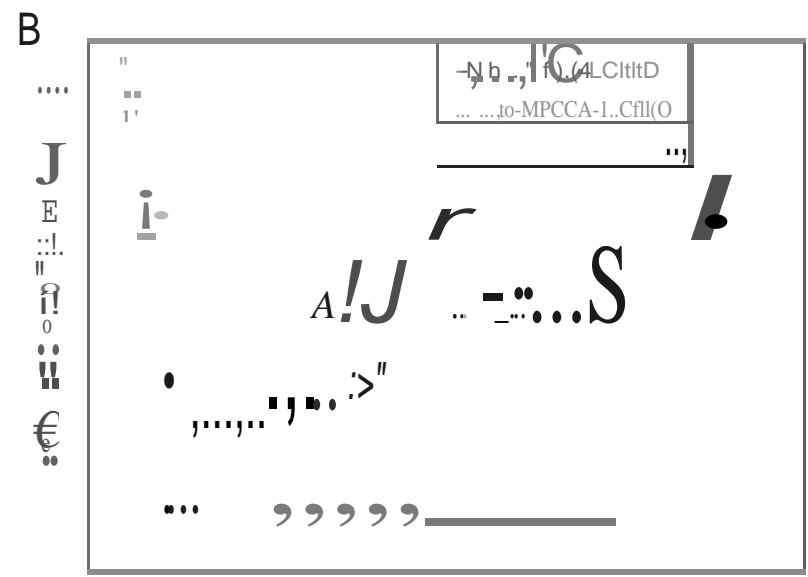

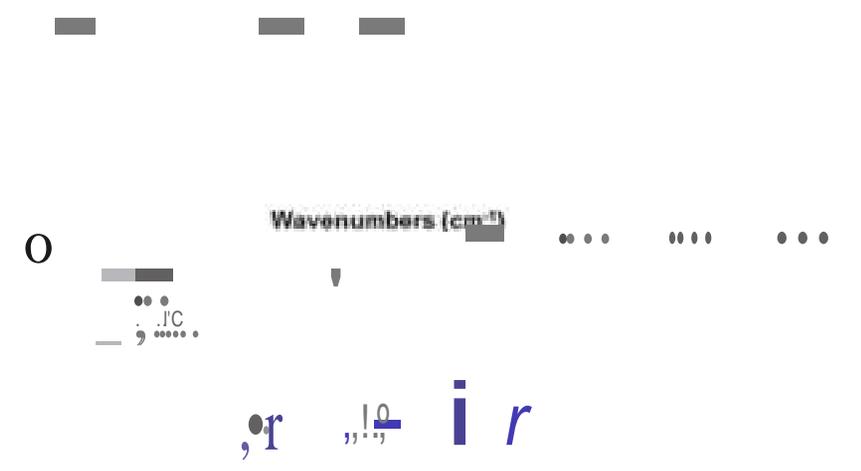

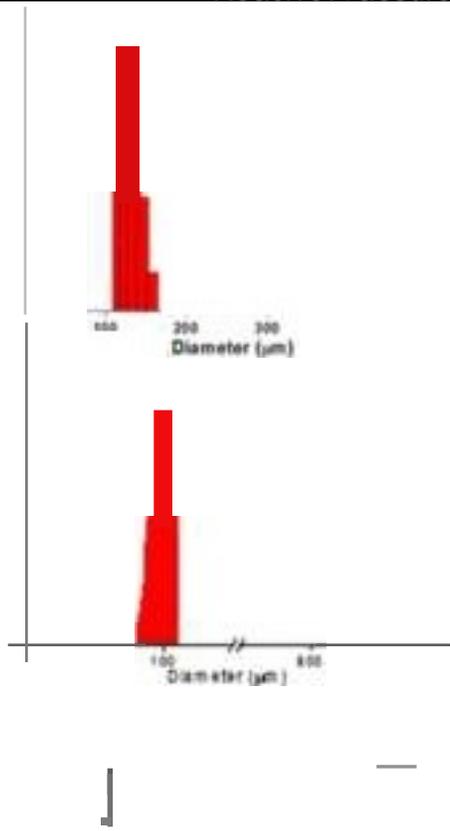
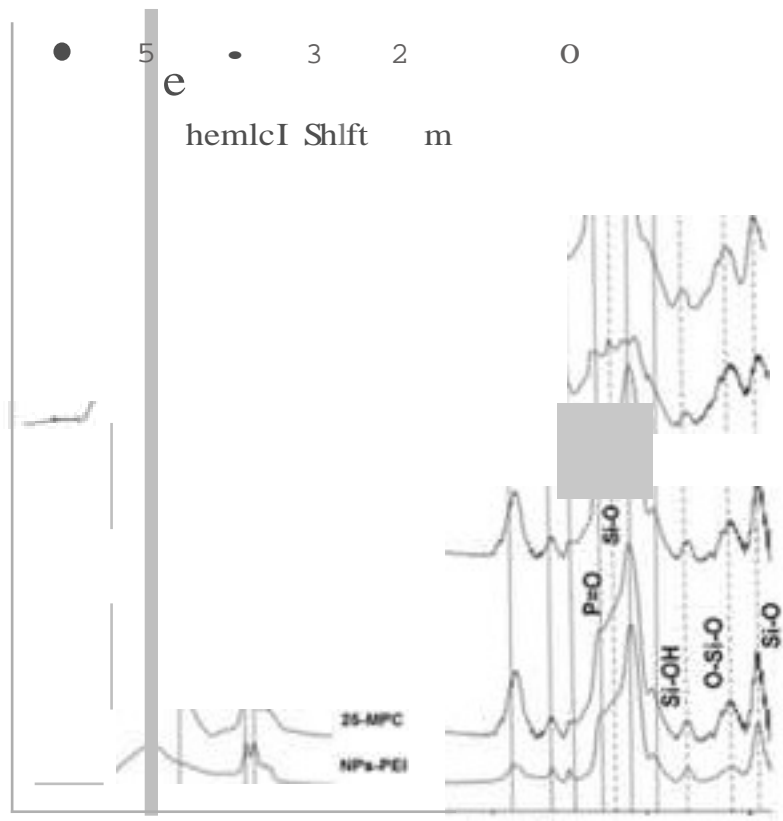
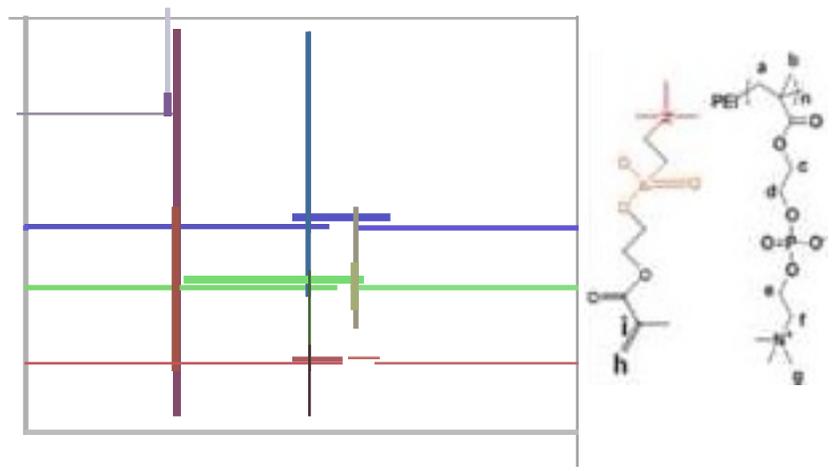

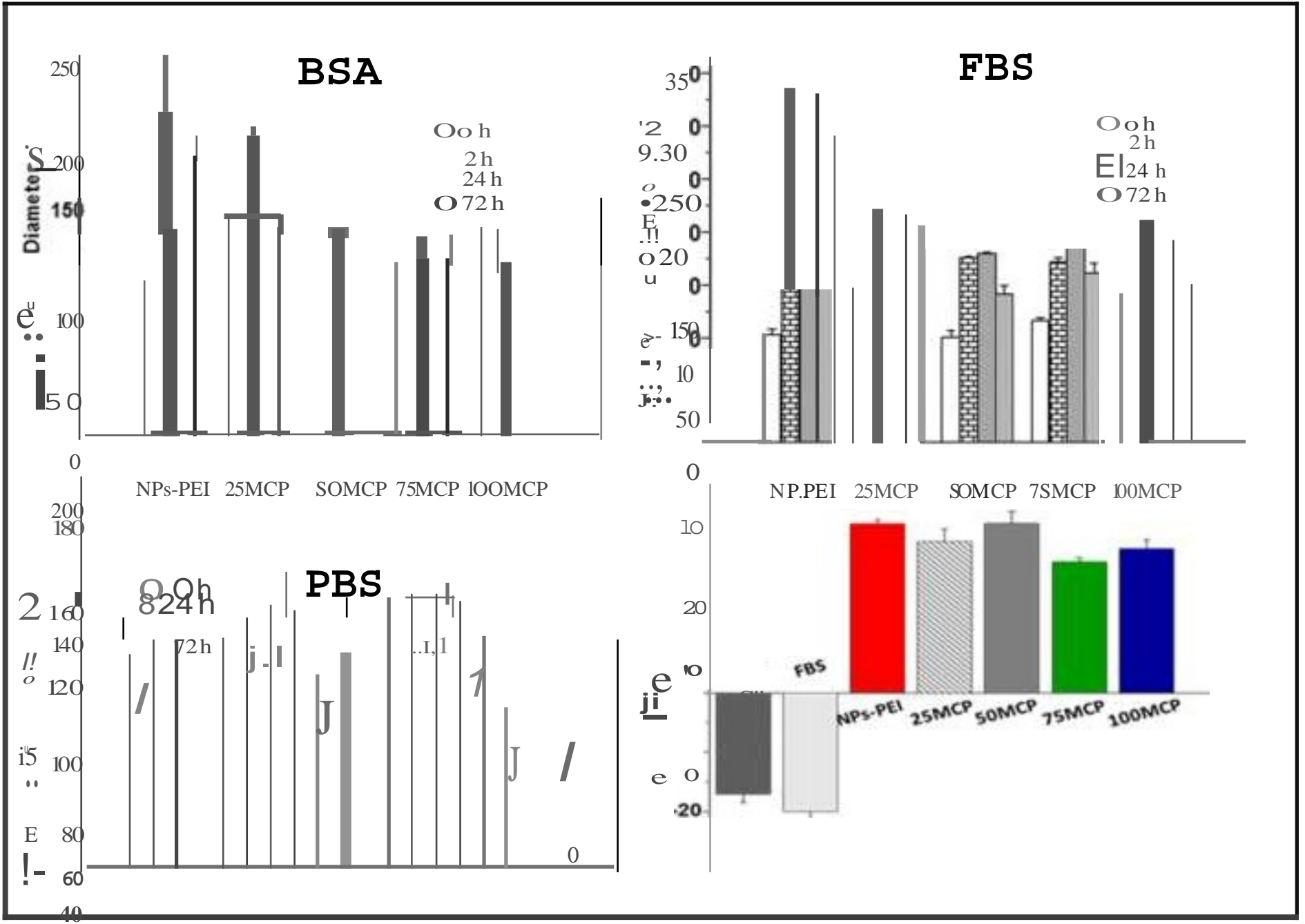

NPs PEI  .!!!
25MCP        O
             D:-.10
             N
50MCP

75MCP
                -30
100MCP

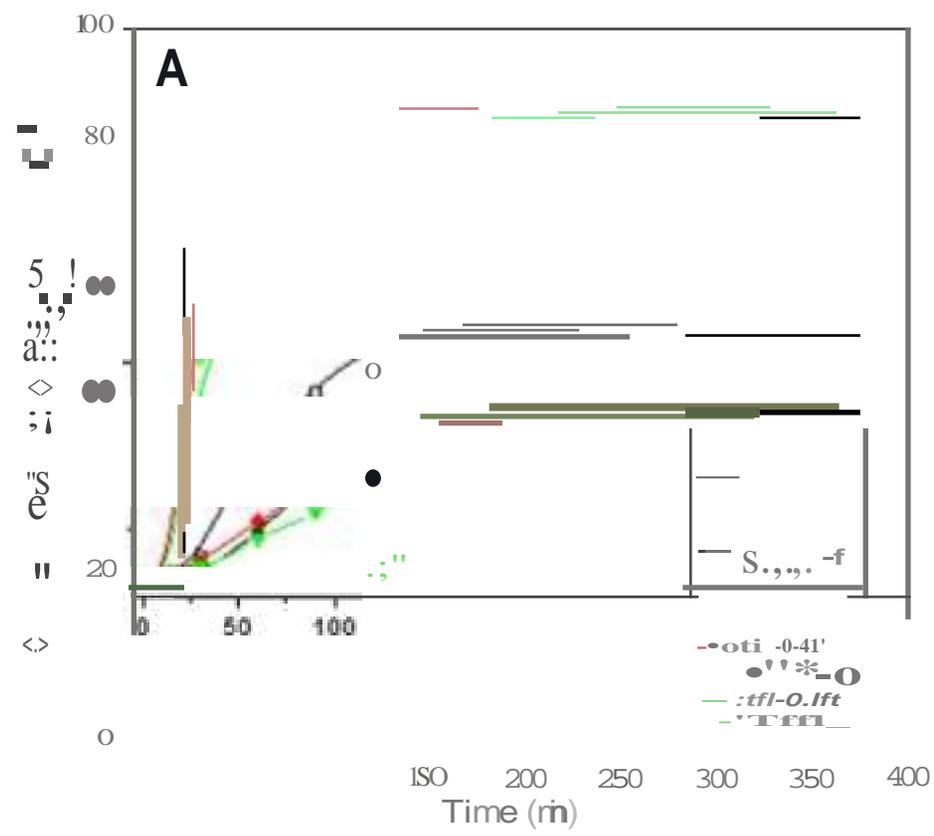

| SAMPLE | FIRST ORDER KINETIC MODEL |
|---|---|
| NPs-PEI-RT | Y = 36(1−e^(−0.0039t))1..2 |
| NPs-PEI-50ºC | Y = 48(1−e^(−0.0<nt t3)) |
| 50-MPC-RT | Y = 38(1−e.∞osi¡t.t) |
| 50-MPC-50ºC | Y = 89(1−e.o.ou i o.s.i) |
| 75-MPC-RT | Y = 35(1−e.O.Ot6t)L2 |
| 75 MPC 50ºC | Y = 87(1−e.∞i0t o&2) |

**Figure 5**
Click here to download high resolution image

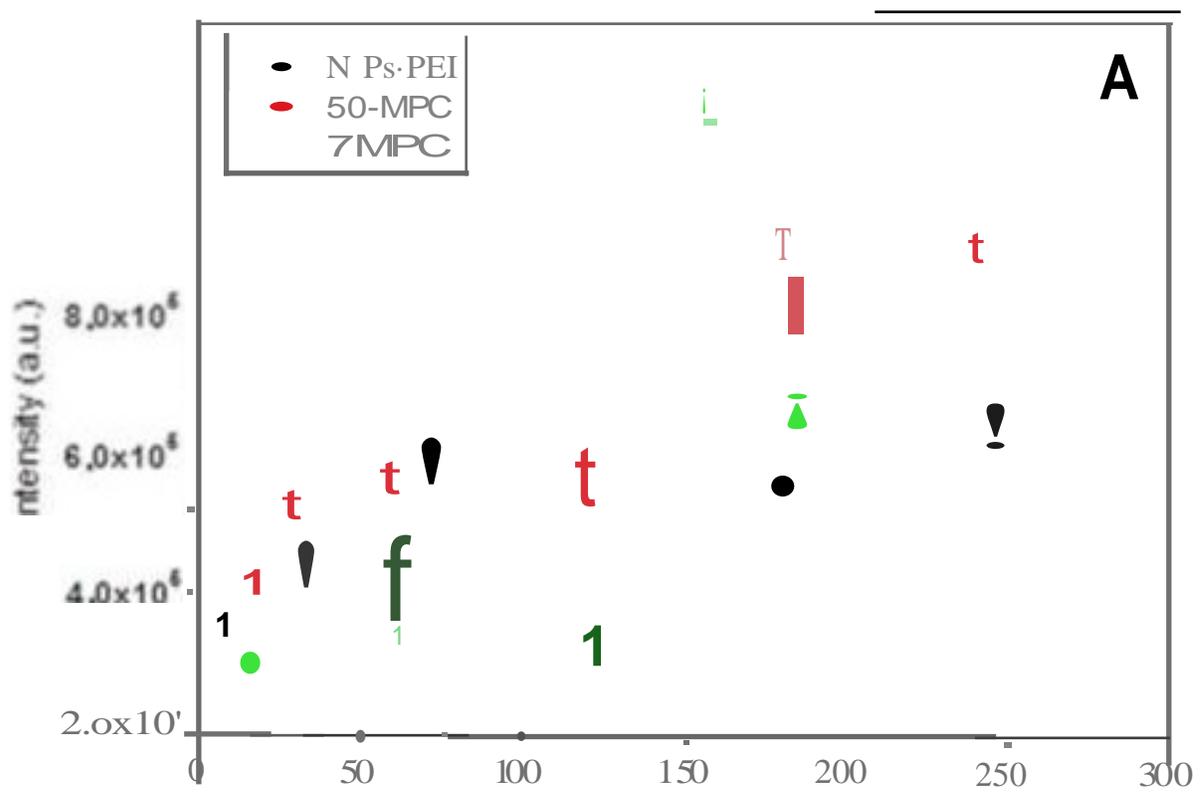

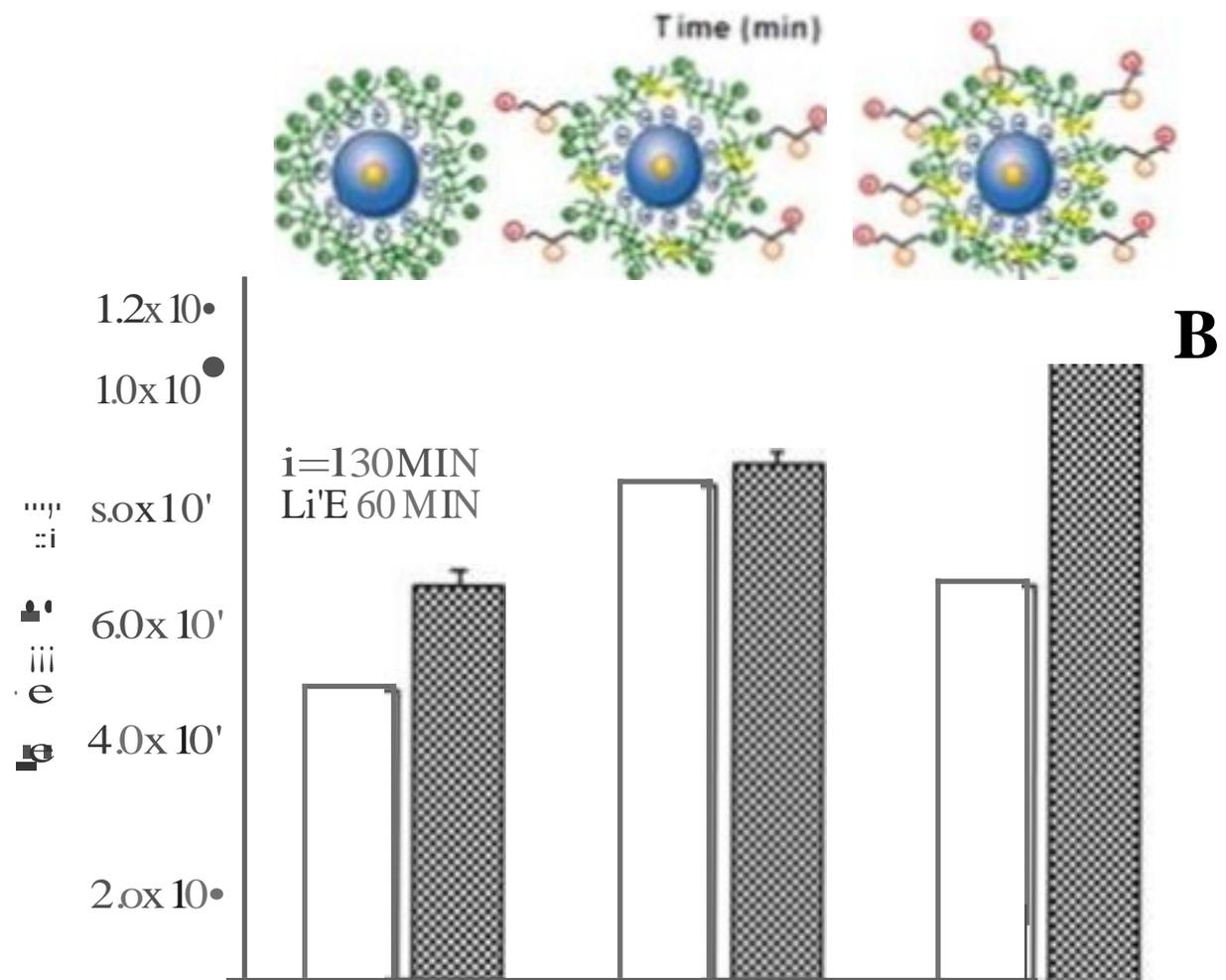

o ──────────────
NPs-PEI     50-MPC     75-MPC

A

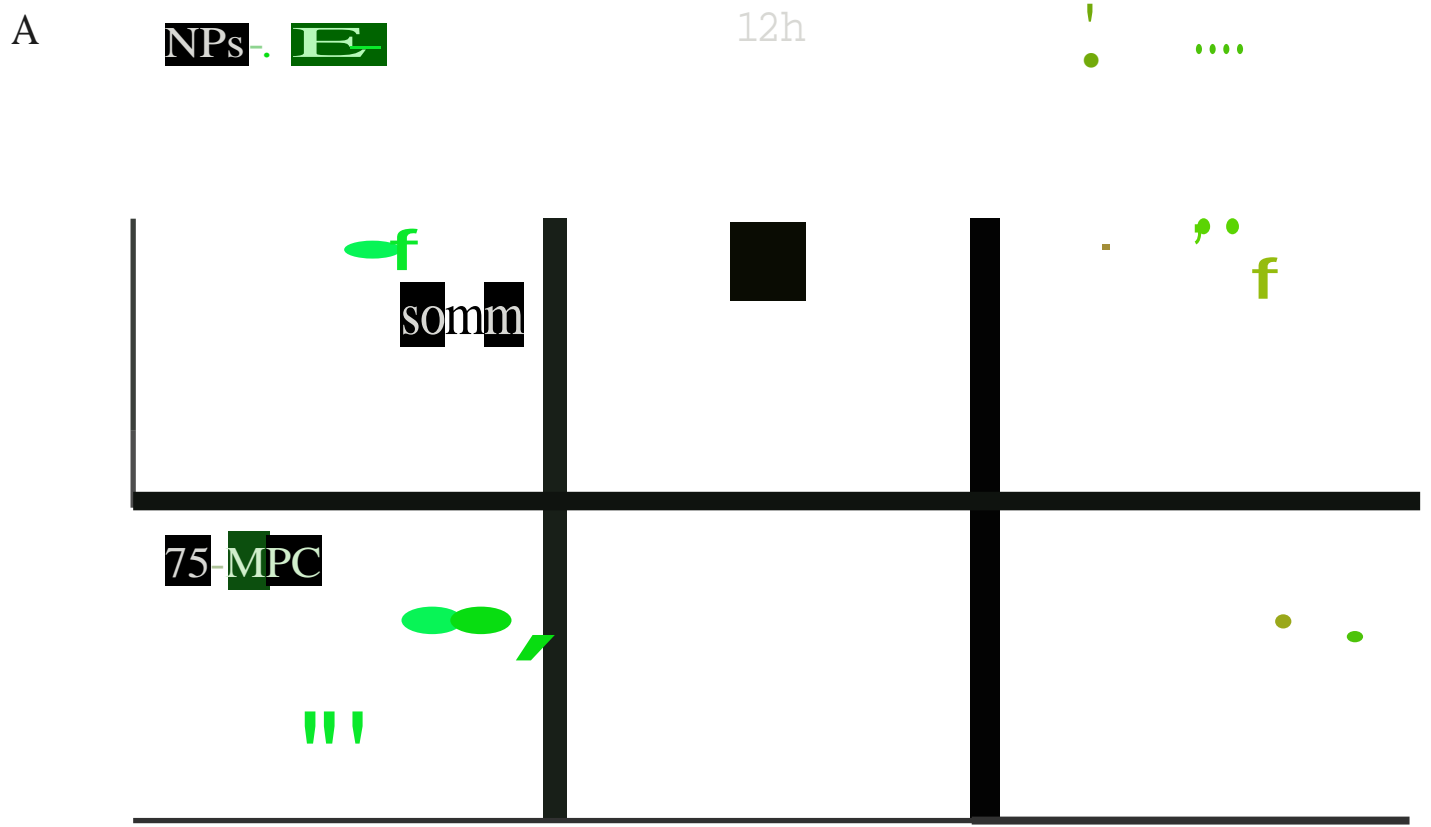

B

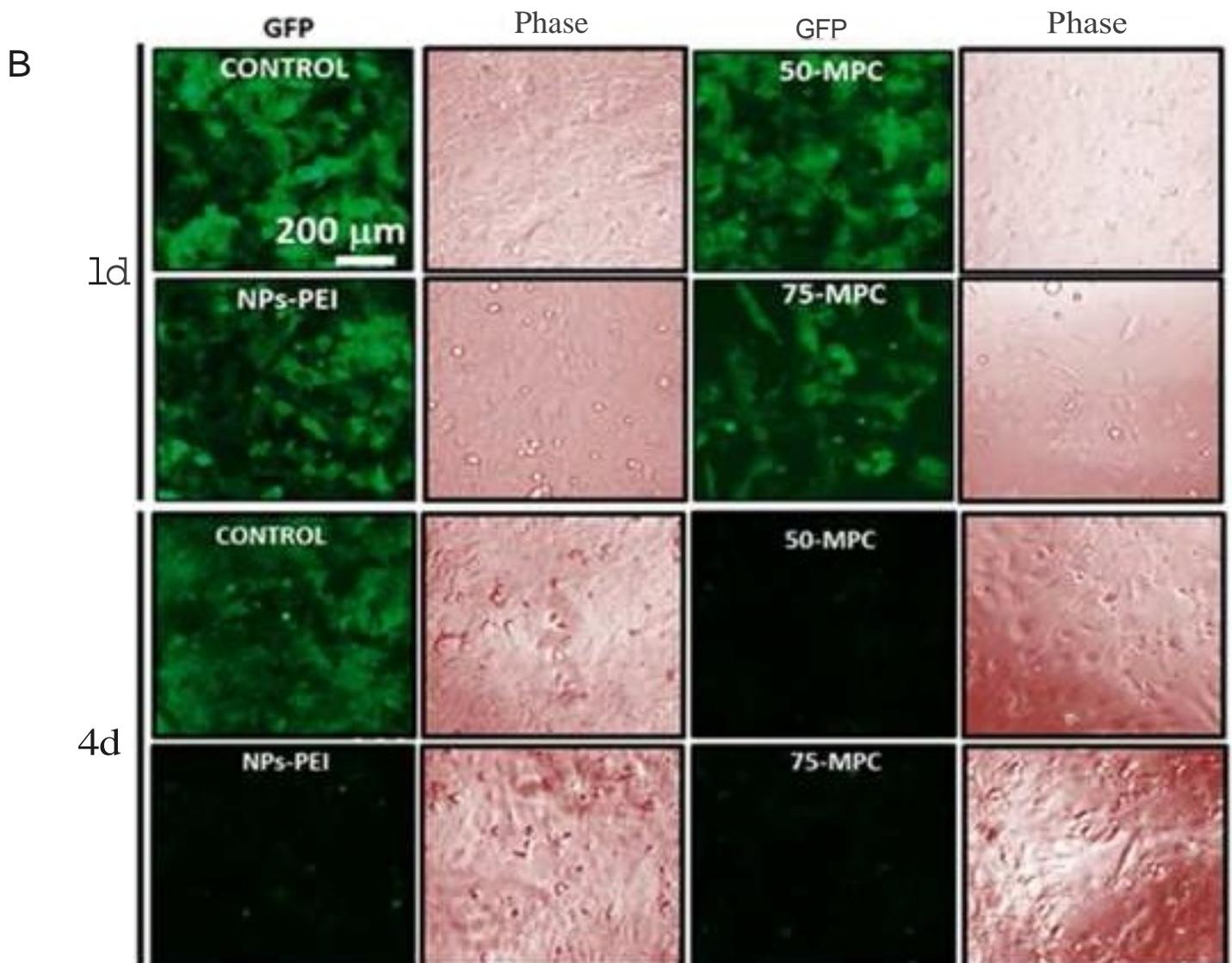

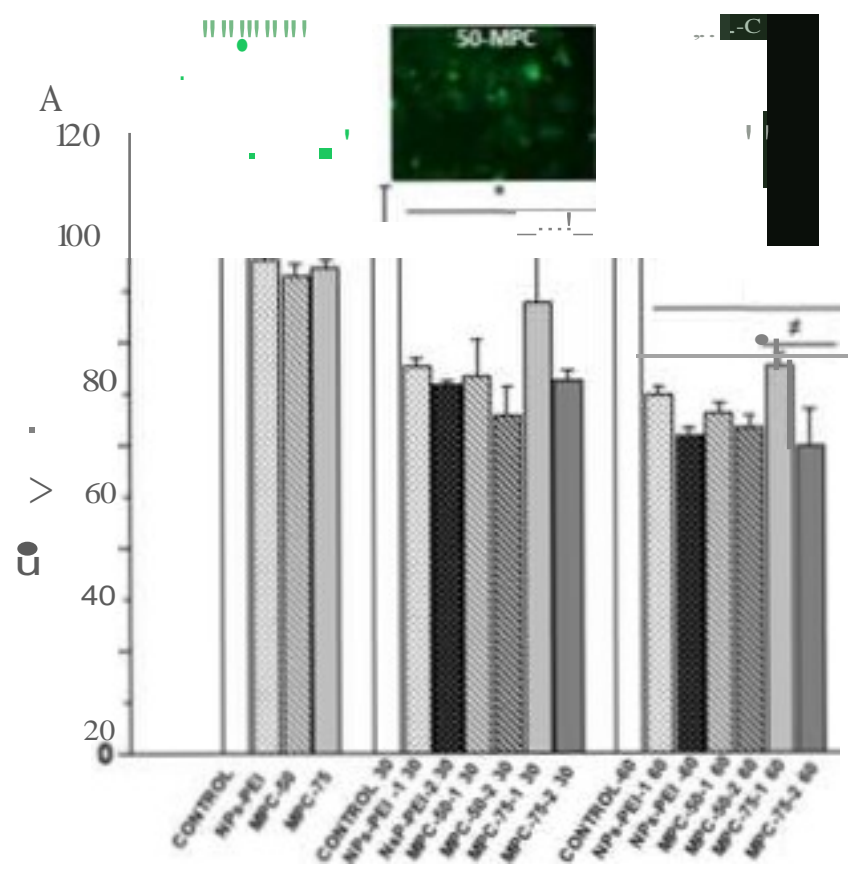
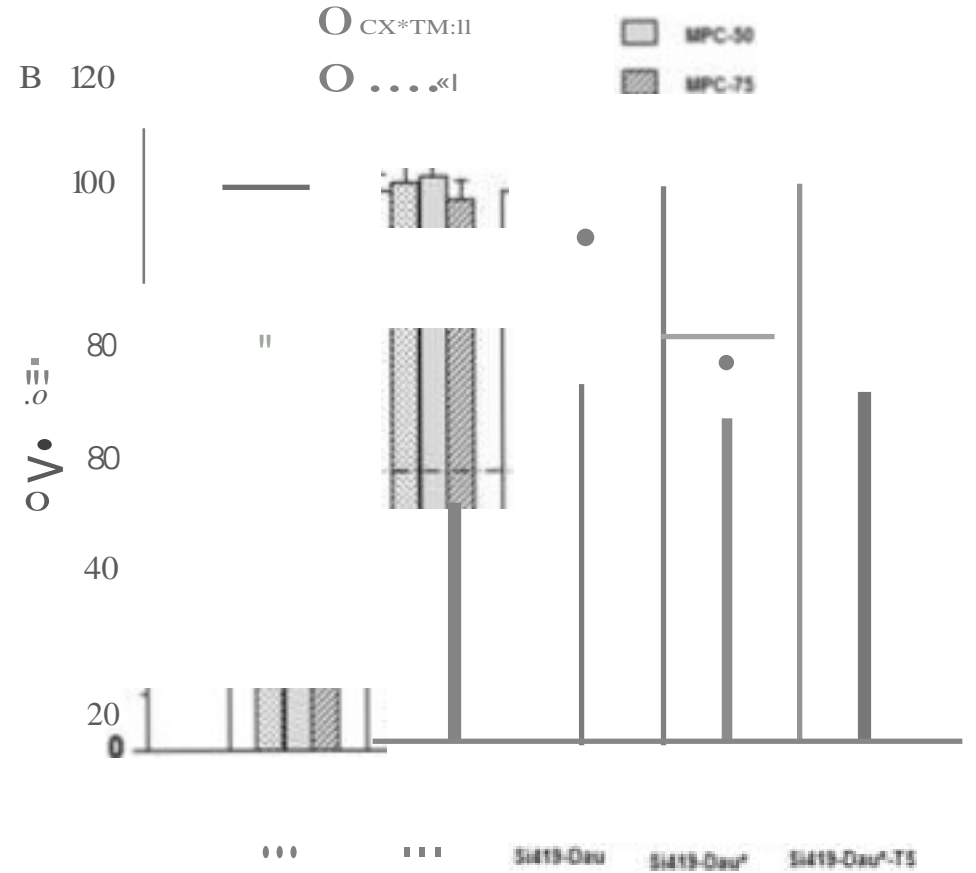

DRUG RELEASE

TWIST-SILENCING